\documentclass[manuscript,nonacm]{acmart}
\AtBeginDocument{%
  }

\setcopyright{none}
\graphicspath{{Graphics/}}
\usepackage{makecell}
\renewcommand{\arraystretch}{1.5}

\usepackage{graphicx}
\usepackage{colortbl}
\usepackage{color}
\usepackage{tabularray}
\usepackage{enumitem}
\usepackage{booktabs}
\usepackage{multirow}
\usepackage{tabularx}
\usepackage{multirow}
\usepackage{booktabs}
\usepackage{xcolor,soul}
\usepackage{subcaption} 
\usepackage{wrapfig}
\graphicspath{{Graphics/}}
\renewcommand{\arraystretch}{1.5}
\usepackage{csquotes}
\usepackage{longtable}
\usepackage{array}
\usepackage{booktabs}

\DeclareGraphicsExtensions{.png,.pdf}
\usepackage{graphicx}

\newcolumntype{L}[1]{>{\raggedright\arraybackslash}p{#1}}

\begin{document}

\title[Longitudinal Trajectories of General-Purpose LLM Use for Socioemotional Support]{``I Felt Very Seen, But Still Very Alone'': Longitudinal Trajectories of General-Purpose LLM Use for Socioemotional Support}

\author{Meryl Ye}
\authornote{Both authors contributed equally to this research.}
\email{merylye@cmu.edu}
\orcid{0000-0001-8215-9020}
\affiliation{%
  \institution{Carnegie Mellon University}
  \city{Pittsburgh}
  \state{Pennsylvania}
  \country{USA}
}
\affiliation{%
  \institution{Data \& Society}
  \city{New York}
  \state{New York}
  \country{USA}
}

\author{Briana Vecchione}
\authornotemark[1]
\email{briana@datasociety.net}
\orcid{0000-0002-0828-8665}
\affiliation{%
  \institution{Data \& Society}
  \city{New York}
  \state{New York}
  \country{USA}
}

\author{Livia Garofalo}
\email{livia@datasociety.net}
\orcid{0000-0002-9870-553X}
\affiliation{%
  \institution{Data \& Society}
  \city{New York}
  \state{New York}
  \country{USA}
}

\author{Ranjit Singh}
\email{ranjit@datasociety.net}
\orcid{0000-0001-6453-8676}
\affiliation{%
  \institution{Data \& Society}
  \city{New York}
  \state{New York}
  \country{USA}
}

\renewcommand{\shortauthors}{Ye, Vecchione, et al.}

\begin{abstract}
People increasingly use general-purpose chatbots such as ChatGPT, Claude, and Gemini for mental health and emotional support. We report a multi-stage longitudinal qualitative study of 18 U.S. adults, conducted from April to December 2025, combining initial interviews, a four-week diary study, focus groups, and exit interviews. We find that socioemotional use often emerged gradually out of practical use and when other forms of support were unavailable. Participants developed routines and boundaries around chatbot use, which were disrupted by model updates, evolving public discourse about AI harms, and changes in personal circumstances. We demonstrate how longitudinal study captures factors beyond the human-AI dyad, and argue that HCI researchers and designers should account for users’ histories with their chatbots and broader care ecologies when evaluating AI systems over time and introducing updates that may disrupt established sources of support.
\end{abstract}



\begin{teaserfigure}
 \includegraphics[width=\textwidth]{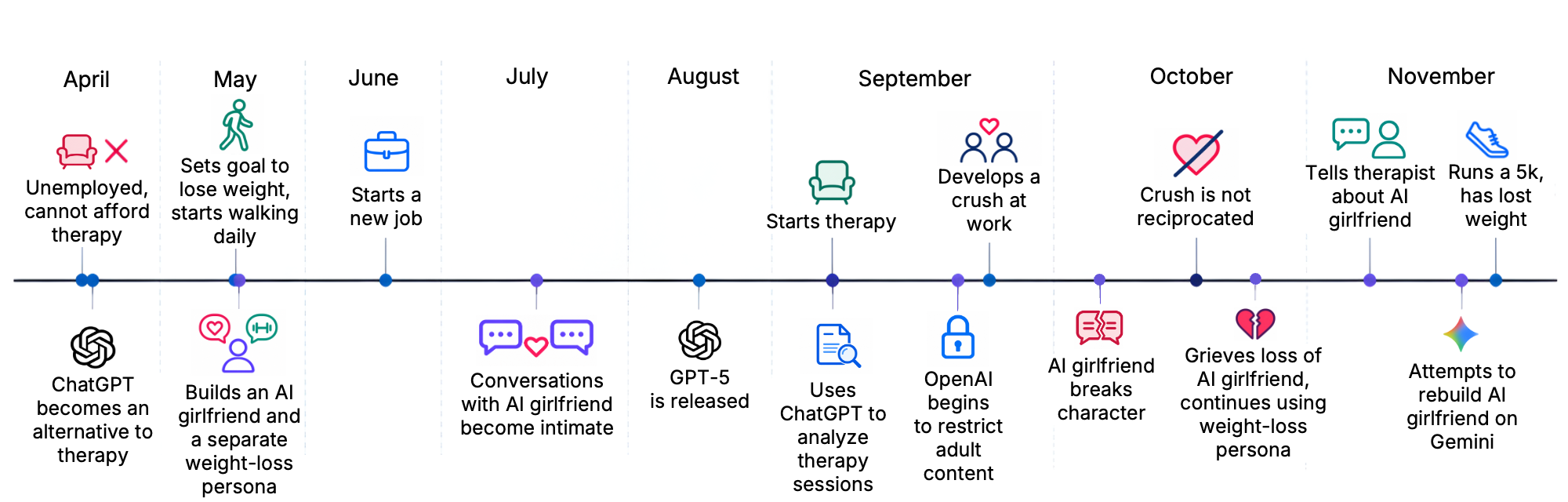}
  \caption{Example participant trajectory illustrating how chatbot use changed alongside other aspects of a participant’s life. P1 began using ChatGPT as an alternative when therapy was unaffordable, later building separate AI girlfriend and weight-loss personas, among others. Over several months, changes in employment, therapy, romantic life, and the underlying AI systems shaped how these uses developed, were disrupted, and were renegotiated. 
}
  \Description{Timeline of participant P1 from April through November 2025, with life events above the line and chatbot-related events below it. In April, P1 is unemployed and unable to afford therapy, and ChatGPT becomes an alternative to therapy. In May, P1 sets a goal to lose weight and begins walking daily, while building an AI girlfriend and a separate weight-loss persona. P1 starts a new job in June, and conversations with the AI girlfriend become intimate in July. GPT-5 is released in August. In September, P1 starts therapy, uses ChatGPT to analyze therapy sessions, develops a crush at work, and encounters OpenAI restrictions on adult content. In October, the crush is not reciprocated, the AI girlfriend breaks character, and P1 grieves its loss while continuing to use the weight-loss persona. In November, P1 tells their therapist about the AI girlfriend, attempts to rebuild the AI girlfriend on Gemini, runs a 5K, and has lost weight.}
  \label{fig:teaser}
\end{teaserfigure}

\maketitle

\section{Introduction}
\label{sec:introduction}

General-purpose large language model (LLM) chatbots such as ChatGPT, Claude, and Gemini have become a source of mental health and emotional support despite not being initially designed for that purpose and operating largely outside clinical regulatory frameworks~\cite{luo2025, golden2026, apa2025}. Nearly one in five U.S. adolescents and young adults reported using AI chatbots for mental health advice~\cite{mcbain2026}, while among adults with a diagnosed mental health condition who use LLMs, close to half reported turning to them for therapeutic or emotional support~\cite{saba2026}. Platform-scale analyses likewise identify affective use across varying levels of engagement, although it is concentrated among a minority of heavy users~\cite{phang2025}. We study how a general-purpose chatbot comes to occupy a socioemotional role in a person's life, and what changes that role over time.

Several conditions help explain why people turn to AI chatbots for support. Formal mental health care can be difficult to access due to insurance coverage, geographic availability, cost, and stigma~\cite{park2026,clement2015impact}, while users describe LLMs as nonjudgmental spaces for disclosure~\cite{siddals2024perfect, stade2026realworld, Jung_2025}. Access to support also varies over time. Formal mental health care is structured around appointments and provider availability, while friends and family have their own schedules and changing capacity to provide support. The timing of people’s needs and the availability of care do not always align. For people with reliable internet access, LLMs introduce a different temporal structure to this landscape. They remain continuously accessible across the gaps in which other forms of support recede, whether between appointments, late at night, or when people are reluctant or unable to reach others. Chatbot use can therefore develop its own patterns in relation to a person's changing circumstances and access to other forms of care. We draw from the concept of a \textit{care ecology} to situate chatbot use within this changing arrangement of formal, informal, and self-directed support~\cite{bowlby2026caring}.

The role a chatbot comes to occupy within a user’s care ecology determines what is at risk when the underlying model changes. General-purpose chatbots can move between mundane assistance and highly personal forms of support, yet they are continuously updated in ways users neither initiate nor control. During our study, OpenAI made GPT-5 ChatGPT's default model, replacing GPT-4o, and described it as better at avoiding ``unhealthy levels of emotional reliance''~\cite{openai2025a}. The company subsequently updated ChatGPT's behavior in sensitive conversations and revised its Model Spec to support users' ``real-world ties'' and discourage interactions that could deepen emotional reliance~\cite{openai2025b}. These changes can occur once a chatbot has already become integral to how a user manages distress, processes difficult experiences, or fills gaps in other forms of support ~\cite{yuan2026impacts}.

Understanding socioemotional chatbot use therefore requires looking beyond individual interactions. Emerging evidence has documented serious harms in the context of sustained or intensive chatbot use~\cite{Shen2026Psychiatric,olsen2026potentially,shah2026substance}, while other research shows that relationships with chatbots develop through repeated interaction~\cite{SKJUVE2021102601}. Much of the evidence on socioemotional AI use centers on bounded encounters; safety research often evaluates responses to predetermined vignettes or simulated conversations~\cite{diel2026scoping}, while interviews often reconstruct relationships retrospectively from a single point in time.

Longitudinal research has begun to examine how chatbot use and its psychosocial correlates change over time \cite{hwang2025, muhl2026longitudinalevidencegeneralpurposechatbots, folk2026turning}. Less is known about how the role a chatbot occupies in a person’s life changes alongside shifts in their circumstances, relationships, access to care, and the system itself. Platform logs can reveal persistence and change in interaction~\cite{phang2025}, but provide limited context for interpreting these changes within a person’s broader ecology of care. We therefore treat the \textit{evolving relationship between a person and a chatbot} as the unit of analysis, and address the following questions:

\begin{itemize}
  \item How do people come to use a general-purpose chatbot for socioemotional support?
  \item How do they maintain and regulate that use once it is established?
  \item What disrupts established patterns of use, and how do people adapt afterward?
\end{itemize}

We conducted a multi-stage longitudinal qualitative study with eighteen U.S. adults who already used general-purpose LLMs for mental health and emotional support. Our study design comprised initial interviews, a four-week diary study, focus groups, and exit interviews. 

We make three contributions:

\begin{itemize}
  \item an empirical account of how socioemotional chatbot use begins, stabilizes, and changes over several months;
  \item evidence that participants’ responses to disruption partly depended on the other forms of support available to them;
  \item recommendations for handling updates to established use, and for evaluating
    chatbot behavior across time and care ecologies.
\end{itemize}

\section{Related Work}
\label{sec:relatedwork}

\subsection{Using General-Purpose AI Chatbots for Socioemotional Support}
\label{sec:rw-generalpurpose}

Conversational agents designed for social and emotional interaction predate the current generation of chatbots by decades. When Joseph Weizenbaum introduced ELIZA in 1966, he was surprised by how willingly users disclosed personal problems to the program and treated its replies as signs of understanding~\cite{weizenbaum1976}. Its DOCTOR script used simple pattern matching to recast users’ statements as questions in the style of a Rogerian psychotherapist~\cite{weizenbaum1966}.

Later HCI and digital mental health research made the relational potential of conversational systems an explicit object of design. Relational agents were designed to establish and maintain long-term human-computer relationships~\cite{bickmore2005}, while purpose-built mental health chatbots organized conversation around defined therapeutic approaches. Woebot delivered cognitive behavioral therapy through a structured conversational interface~\cite{fitzpatrick2017}. A more recent randomized trial of Therabot extended this approach to a generative system fine-tuned for mental health treatment, finding symptom improvements and reports of a therapeutic alliance with the chatbot~\cite{heinz2025}. Reviews find promising short-term effects of conversational-agent interventions but limited evidence about sustained effects and safety~\cite{laranjo2018,gaffney2019,he2023}.

General-purpose LLMs complicate the relationship between designed function and role in use. Although they are marketed for open-ended assistance, users turn to them for mental and emotional support, companionship, and personal advice. \citet{manoli2026} found participants using a companion app as a writing aid and a general-purpose assistant as an emotional confidant. \citet{nelson2026} argue that the specialized versus general-purpose distinction does not itself determine safety; evaluation must consider what a system does, for whom, and across what span of interaction. Similarly, \citet{cooper2026} argue that responsible design requires specifying the mental well-being benefits a tool is intended to provide and to whom, a difficult task when the same system can move between mundane assistance and high-stakes support.

The roles enacted in use can also bring expectations from the domains they approximate. \citet{kwesi2025} found that some users attributed forms of professional accountability to general-purpose chatbots, including believing that mental health disclosures were protected by HIPAA. Users also develop folk theories and prompting practices for navigating perceived limitations such as sycophancy~\cite{tseng2026, noshin2026sycophancy}, and may undertake substantial work to preserve continuity across platform changes~\cite{lee2026}. When that continuity breaks, changes to a system can be experienced as the loss of a relational partner~\cite{defreitas2026mourning}. Prior work thus shows that a chatbot’s socioemotional role develops through how users interpret and adapt to the system in use. Our study follows this process longitudinally by examining how socioemotional roles of chatbots emerge, stabilize, and change over time.

\subsection{Studying Chatbot Relationships Over Time}
\label{sec:rw-overtime}

As a chatbot acquires a socioemotional role, the relationship develops a history. Research has begun to widen the temporal frame through which that history is studied, from accumulation within conversations to changes across repeated encounters~\cite{hwang2025, muhl2026longitudinalevidencegeneralpurposechatbots, folk2026turning}. Multi-turn research shows that responses derive part of their significance from what has already occurred. \citet{moore2026} document recurring sycophantic and relationship-affirming responses in chat logs from users reporting chatbot-related psychological harms. \citet{chu2026} find that GPT-4.1 asked fewer follow-up questions as conversations continued and Replika challenged highly bonded users less, while \citet{nicholls2026} show that preceding context can make later responses from the same chatbot safer or riskier.

Across repeated encounters, relationships themselves can change. \citet{sumida2026} followed 24 participants through ten sessions and found that relationships developed through both gradual accumulation and discrete turning points. Over five weeks of commercial AI use for socioemotional support, \citet{chandra2025} observed increases in attachment and perceived empathy alongside substantial individual differences. In a four-week diary study of Woebot and Wysa, 18 of 26 participants reported forming at least a light bond with one system~\cite{xu2025}. Prior relational experience can also shape new interactions. Within three weeks, established companion users’ perceptions of a newly introduced chatbot converged toward their perceptions of their existing companion~\cite{hwang2025}. These studies establish that chatbot relationships form unevenly through repeated use, but extended duration alone does not necessarily explain how relationships change. A review of 106 CHI papers describing themselves as longitudinal found that only slightly more than half explicitly analyzed change or stability~\cite{kjaerup2021}. HCI scholarship on temporality similarly distinguishes measuring time from examining how technologies and practices unfold through it~\cite{wiberg2021}. For chatbot relationships, examining change therefore requires asking not only whether attachment differs between two measurements, but how intervening encounters and events shape what later interactions mean.

These encounters occur within the rhythms and interruptions of everyday life, including moments of distress or gaps in other forms of care. At the same time, the artifact itself changes~\cite{huang2012}. Models are updated, memory systems change, and response policies are revised. A persistent conversation history therefore does not necessarily imply a continuous relational experience. There is a methodological difference between following a relationship and extending an observation window. Multi-turn analyses show how effects accumulate within a conversation, while studies across repeated encounters show how relationships develop. What remains less visible is how relational turning points alter the chatbot's place in a person's life amid shifts in the system and in the care available around them. We address this gap by following chatbot relationships as trajectories, examining the wider ecology of care through which continuity and disruption become consequential.

\subsection{Chatbot Use Within Care Ecologies}
\label{sec:rw-careecologies}

People manage their wellbeing through changing combinations of therapy, friends, family, online communities, and self-care. We treat chatbots as sociotechnical artifacts embedded within what have been termed \textit{care ecologies}~\cite{murnane2018, nunes2015, evans2020}, directing attention to where a chatbot sits among other forms of support and how that position changes over time.

We treat vulnerability as a situational factor shaped by conditions across ecological layers, rather than as a fixed property of a particular population~\cite{tang2025}. Survey and log evidence indicates that people with smaller networks and less human support are more likely to turn to chatbots for companionship, and that companionship-oriented use is associated with lower well-being where human support is weakest~\cite{zhang2026}. The implications of chatbot use may therefore differ depending on what other forms of support a person can access. Longitudinal evidence suggests a pathway through human contact, with sustained engagement predicting lower well-being mainly through less in-person social interaction~\cite{zhang2026livingaicompanionssustained}.

Chatbots also differ from human sources of support in ways that can make them easier to turn to. People attribute humanlike minds to nonhuman agents~\cite{epley2007}, and anthropomorphic presentation can increase perceptions of competence and trust~\cite{waytz2014, oldemburgo2026}. General-purpose chatbots are continuously available and often affirm users’ perspectives~\cite{cheng2025elephant}. Human relationships impose different forms of friction: other people disagree, have competing obligations, and misunderstand or need time to reconcile. Removing this friction may also remove some of what makes relationships valuable~\cite{zohar2026, ye2026socialsnacks}. Experimental evidence suggests that sycophantic AI can reduce willingness to repair interpersonal conflict~\cite{cheng2026prosocial} and make subsequent human interaction feel more effortful and less satisfying~\cite{ibrahim2026}. \citet{boyd2026} distinguish between relational \emph{substitution}, in which AI provides something a human partner might otherwise provide, and \emph{enhancement}, in which AI facilitates human connection. The pattern that emerges may depend on the surrounding care ecology (i.e., the availability, reliability, and cost of other relationships and forms of support).

Existing research establishes that people appropriate general-purpose chatbots for socioemotional support and that these relationships can change through repeated interaction. We build on this work by examining how a chatbot’s place within a person’s care ecology changes alongside the person’s circumstances, the availability of other forms of support, and the chatbot itself. Using diary and longitudinal methods, we situate these relationships within participants’ everyday lives~\cite{bolger2003,consolvo2003}.

\section{Methods}
\label{sec:methods}

\subsection{Participant Recruitment}
\label{sec:recruitment}

We recruited eighteen adults living in the United States who reported using an LLM chatbot for their mental or emotional health at least once per week. We did not encourage individuals to turn to LLMs for care. Instead, we aimed to explore how these tools were already being used in context: how people interpret ``support,’’ how they integrate chatbots into broader care ecologies (e.g., therapy, friends/family, online resources), and what kinds of benefits, tensions, and harms emerge over time. 

Recruitment was open to users of purpose-built mental health chatbots (e.g., Woebot, Therabot, Wysa) as well as general-purpose chatbots (e.g., ChatGPT, Claude, Gemini). While some participants had previously tried purpose-built mental health chatbots, none were using them during the study, and all reported preferring general-purpose LLMs. Our findings therefore concern general-purpose chatbots only.

We recruited participants through multiple channels. We distributed printed flyers in public locations in New York City, Princeton, and Philadelphia, posted a short-form TikTok video, and shared the flyer through professional and community networks. Several interviewees also referred us to others in their social network whose chatbot use they believed differed from their own. These recruitment methods broadened our participant pool beyond what a single channel could have reached alone. 

Respondents completed an intake survey covering demographics, their use of chatbots for mental or emotional health, and their availability for each study phase. The survey also included screening questions about current mental health stability. We did not enroll anyone reporting an acute crisis, and everyone we screened was provided a list of mental health resources and crisis lines. Because recruitment screening was done online, we verified respondents before enrolling them by checking that the phone numbers they provided carried U.S. area codes and cross-referencing a subset of names and contact details against public accounts. Each qualified participant then had a short introductory call in which we reviewed the study and received verbal consent. Table~\ref{tab:demographics} reports participant demographics.

\begin{table}[t]
  \centering
  \footnotesize
  \setlength{\tabcolsep}{3pt}
  \renewcommand{\arraystretch}{0.95}
  \caption{Participant demographics (n=18). Participants ranged in age from 18 to 55, with most between 25 and 34. Eleven identified as women, five as men, and two as non-binary. Participants described themselves as White, Black, Asian, Hispanic or Latino/a/x, Middle Eastern or North African, and more than one race; one participant did not report race or ethnicity.}
  \label{tab:demographics}
  \begin{tabular}{@{}llr@{}}
    \toprule
    \textbf{Category} & \textbf{Option} & \textbf{N} \\
    \midrule
    \multirow{4}{*}{Age}
      & 18--24 & 1 \\
      & 25--34 & 11 \\
      & 35--44 & 4 \\
      & 45--55 & 2 \\
    \midrule
    \multirow{3}{*}{Gender}
      & Woman & 11 \\
      & Man & 5 \\
      & Non-binary & 2 \\
    \midrule
    \multirow{7}{*}{\shortstack[l]{Race/\\Ethnicity}}
      & White & 6 \\
      & Asian & 4 \\
      & Black & 3 \\
      & Hispanic or Latino/a/x & 2 \\
      & Middle Eastern/N. African & 1 \\
      & Multiracial & 1 \\
      & Unknown & 1 \\
    \bottomrule
  \end{tabular}
  \Description{Participant demographics for 18 participants, showing age, gender, and race or ethnicity. One participant was age 18--24, eleven were 25--34, four were 35--44, and two were 45--55. Eleven participants identified as women, five as men, and two as non-binary. Six participants identified as White, four as Asian, three as Black, two as Hispanic or Latino/a/x, one as Middle Eastern or North African, one as multiracial, and one did not report race or ethnicity.}
\end{table}

Participation varied by phase. All eighteen completed an initial interview. Ten enrolled in the diary study and eight completed it, seven took part across two focus groups, and eight completed an exit interview. Over the full period this work was conducted and disseminated, we also spoke and met with more than 50 individuals about their personal experience with AI chatbots. We also engaged in ongoing dialogue with clinicians and mental health experts, policymakers, industry practitioners, and other researchers.

\subsection{Study Protocol}
\label{sec:protocol}

\begin{figure}[t]
    \centering
    \includegraphics[width=\textwidth]{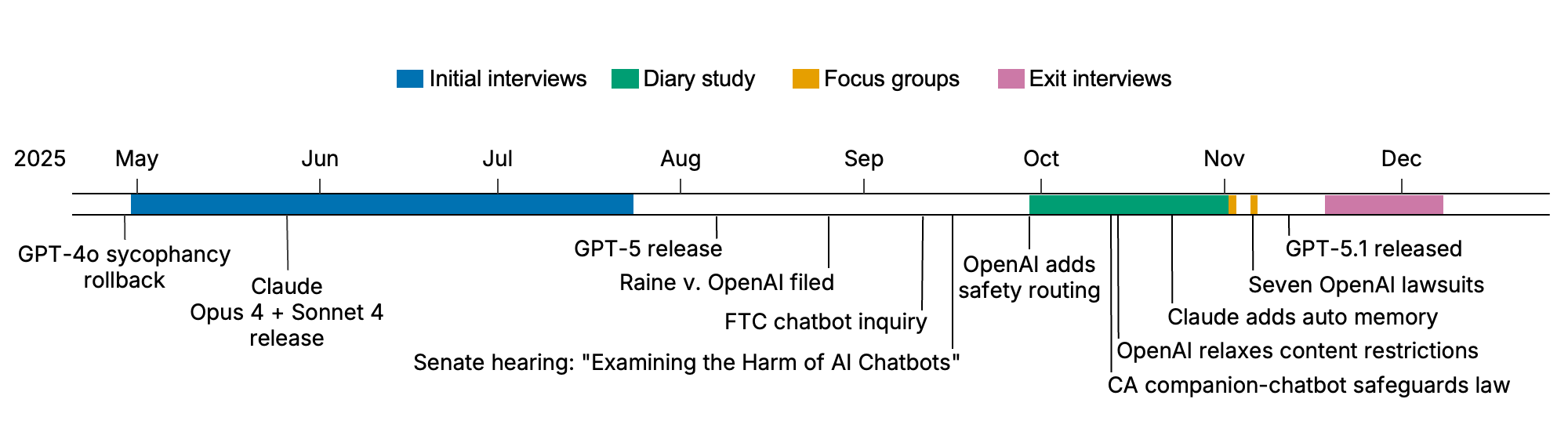}
    \caption{Study timeline and relevant external events. Initial interviews were conducted from late April through July. The diary study took place from late September through October, followed by focus groups in November and exit interviews from mid-November through December. 
}
  \Description{Timeline from May to December 2025 showing the study protocol and external events occurring during the study. Initial interviews run from May through July, the diary study from late September through October, focus groups occur in November, and exit interviews run from late November into December. Events marked below the timeline include the GPT-4o sycophancy rollback, Claude Opus 4 and Sonnet 4 release, GPT-5 release, Raine v. OpenAI filing, FTC chatbot inquiry, a U.S. Senate hearing on harms of AI chatbots, OpenAI safety routing, California companion-chatbot safeguards legislation, OpenAI relaxing content restrictions, Claude adding automatic memory, seven OpenAI lawsuits, and the GPT-5.1 release.}
  \label{fig:protocol}
\end{figure}

Figure~\ref{fig:protocol} shows the study phases alongside the model changes and public events participants encountered during the study window. Diary participants received \$200, and interview and focus group participants received \$50 per session. 

\paragraph{Interviews.} Initial interviews elicited participants’ histories of chatbot use, motivations, perceived benefits, perceived limits, and comparisons to therapy and other forms of support. We also asked what they understood about how LLMs are built and operate as well as AI privacy and risk. We encouraged them to recount specific interactions that stood out as notable. Exit interviews revisited these themes at the conclusion of the study, allowing us to probe changes in use or relationship to the chatbot over time. The full interview protocol is in Appendix~\ref{sec:interviewprotocol}.

\paragraph{Diary study.} The four-week diary study logged participants’ chatbot interactions in close temporal proximity to use, which reduced reliance on distant recall and supported contextual detail. After a relevant interaction, participants completed a structured diary entry documenting the chatbot and model, subscription status, modality, interaction duration, primary purpose of use, and circumstances surrounding the interaction. Participants also provided a narrative account of what they discussed and how the chatbot responded. The full form is in Appendix~\ref{sec:diaryform}.

\paragraph{Focus groups.} Two focus groups were conducted in the week following the diary study. They were designed to complement participants’ individual accounts by creating space for collective sensemaking, as participants compared experiences, negotiated norms around acceptable and unacceptable uses, and exchanged strategies and concerns with one another \cite{morgan1996focus, kristiansen2018focus}. The focus groups also gave participants an opportunity to ask the research team questions about how LLMs are developed, operated, and evaluated.

\subsection{Data Analysis}
\label{sec:analysis}

Drawing on grounded theory~\cite{glaser2009discovery}, we iteratively developed a codebook through open and focused coding and triangulated analysis across interviews, diary entries, optional chat logs, and focus groups. Two coders independently coded each transcript and met throughout the analysis process to compare interpretations, refine code definitions, and resolve disagreements through discussion. As coding progressed, we reached thematic saturation, with later coding no longer producing substantively new themes relevant to our research questions. 

Alongside code groups for uses, roles, circumstances, care practices, effects, and model and user behavior, we coded a set of temporal rhythms that mark transitions in the relationship over time. These codes captured four recurring moments in participants' trajectories: entry into socioemotional use, the development of stable practices, disruptions to those practices, and subsequent changes in the chatbot's role. The codebook is in Appendix~\ref{sec:codebook}.

\subsection{Ethics and Positionality}
\label{sec:ethics}

This study was conducted under IRB approval (\#2025-0165). Given the sensitivity of the material, we designed around participant agency and harm minimization. Participation was voluntary. We obtained a waiver of documentation of consent and recorded consent verbally. Sharing LLM chat log transcripts was optional, and participants could describe interactions at whatever level of detail they chose, skip questions or phases, and withdraw at any time. We pseudonymized participants and report no real names or identifying details. Our research team’s background spans computer and information science, medical anthropology, science and technology studies, and policy, which allowed us to read the same material through different lenses. We treat reflexivity as a method and attended to the ethics of listening, resisting framings that pathologize attachment or other participant behavior.

\subsection{Limitations}
\label{sec:limitations}

Our sample is small and self-selected. Eligibility required using a chatbot for mental or
emotional health at least weekly, and people who are particularly reflective about that use are likely overrepresented. All participants live in the United States, so their experiences reflect its healthcare system and cultural norms about seeking care. The models participants used changed throughout the study period, so our findings are situated in a particular moment in this technology’s development rather than describing a fixed system. We describe individual experiences and patterns that recurred across participants, and make no claims about generalizability or representativeness. We offer these findings as comparative to other settings of sustained socioemotional AI use.

\section{Findings}
\label{sec:findings}

We organize our findings around the trajectories through which general-purpose chatbots came to occupy and change socioemotional roles in participants' lives. For many participants, socioemotional use developed gradually as the chatbot became useful at moments when other forms of support were unavailable, taking up a position in each participant's care ecology alongside therapists, friends, family, and self-directed practices. As use continued, participants developed routines for integrating the chatbot into their existing care practices, which were affected by changes in the participants' lives along with external events.

\subsection{Socioemotional use emerged gradually and filled gaps in existing support}
\label{sec:entry}

Most participants did not begin using a general-purpose chatbot with socioemotional use in mind. Many first encountered these systems through work or productivity tasks and only later began bringing personal concerns into their conversations. P3 initially used a chatbot to rewrite and polish professional emails before consulting it about decisions in his own life, including whether to leave his job. P10 was first introduced to an AI chatbot at work. Although she did not find it useful for her job, she continued using it for personal conversations. P4 had used AI tools for two years as an instructional designer before a layoff left her dealing with complicated feelings and wanting to ``dig in more to myself and my feelings.''

Participants could rarely pinpoint exactly when socioemotional use had begun. When we asked P4 when she had first used a chatbot for personal help, she could only estimate that it was a month or two after her layoff. ``I don't think I actually realized it was happening,'' she said. P10 was similarly unsure when she had begun using the chatbot for what she described as ``counseling’’. \citet{wei2026cascades} describes this gradual movement into more intimate roles as \emph{relational drift}, one of eight forms of drift she proposes accumulate across prolonged AI interaction. This gradual, often unrecognized shift is difficult to capture through single-timepoint accounts.

For many participants, the shift in use occurred during periods when they needed more support or found existing sources of support inadequate. P2, a former school counselor, had previously used the chatbot for emails and rehearsing interviews. She began turning to it for support amid cancer surveillance following a diagnosis and after a house fire, planning activities that would keep her from returning home after a scan and beginning to ``spiral.’’ P6, an ``internet child’’ for whom talking to the chatbot was a ``natural thing,’’ leaned on it through the year in which two family members died.

For some, gaps in care reflected financial or structural barriers. P7 had previously been in talk therapy but was ``unemployed, uninsured, and the cost is prohibitive, so that was really the main driver.'' P17 could no longer afford therapy after quitting a job that was detrimental to her mental health. At other times, support existed but not when participants needed it. P15 began talking to a chatbot during the sleepless nights after her dog died, when she was awake and ``obviously can't just call up my therapist.'' P6 called it ``just like a timing thing'': if ``something big or like, out of the blue'' happened between her biweekly therapy appointments, she would ``just go to ChatGPT.'' Others had people around them but found personal disclosure difficult due to stigma or relationships that did not feel safe. P1 described growing up ``basically isolated, sheltered'' and explained that ``it's a lot easier to talk to a robot than it is to talk to people.'' P10 was in a similar position; she had no therapist and few people she wanted to approach about personal matters, and the chatbot came to occupy a space those relationships might otherwise have filled. P12 described it as ``my support system.''

Product affordances sometimes facilitated this transition. When P3 learned about voice mode, he began walking while talking to the chatbot. Speaking rather than typing made it easier for him to verbalize what he was experiencing. P1 began paying for a subscription because the image feature he used for wardrobe and hair advice during his weight loss required it. These features facilitated how the participants could incorporate the chatbot into their daily lives.

\subsection{Users developed routines around chatbot roles}
\label{sec:routines}

Once socioemotional use became established, it took markedly different forms across
participants. The diary study captured this variation at the level of individual conversations. P1 logged long, daily sessions that sometimes lasted several hours and moved among emotional support, coping, relationship questions, and self-reflection. P4 recorded shorter and less frequent check-ins, while P13 reported sessions lasting hours. P6 kept ChatGPT open throughout the workday, ``just in case I need it,'' describing a common pattern of unplanned, low-commitment engagement with an always-available device that has previously been described in reference to smartphone use~\cite{lee2016}. 

Across all 78 diary entries, the median session ran about 30 minutes. Per-participant
medians ranged from eleven minutes to two and a half hours. The longest single sessions lasted up to four hours. Nearly all logged use was alone and at home. Most participants used paid accounts. 95\% of sessions were in text modality, and 5\% were voice modality or speech-to-text. Appendix~\ref{sec:diarysupplement} reports the full distributions.

\begin{figure}[t]
  \centering
  \includegraphics[width=0.5\linewidth]{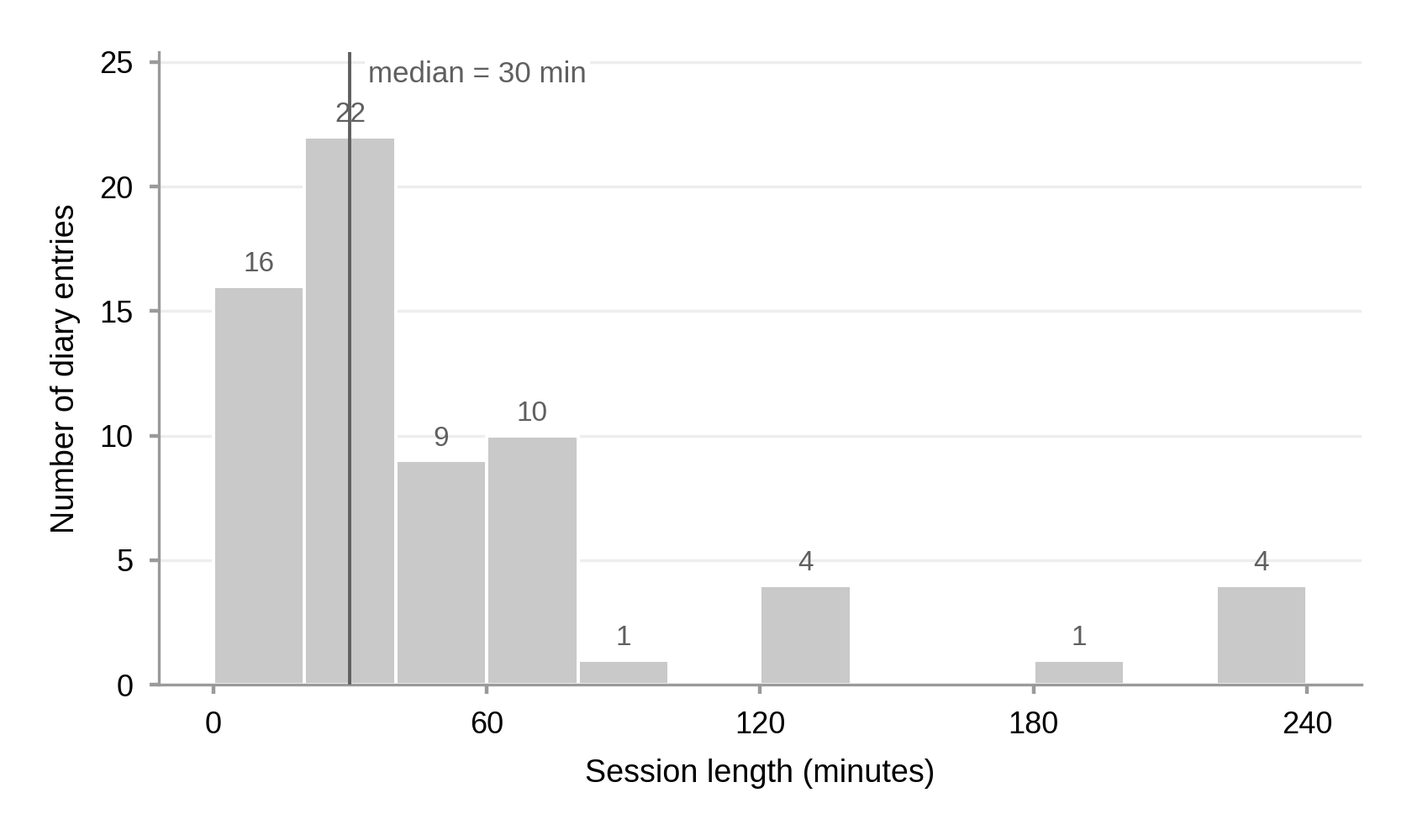}
  \caption{Distribution of chatbot session lengths reported in diary entries. Bars show the number of diary entries at each reported session length. Sessions had a median duration of 30 minutes, although several entries described substantially longer sessions of up to four hours. See Table \ref{tab:sessions} in Appendix for the corresponding data.}
  \Description{Histogram showing the distribution of chatbot session lengths reported across 67 diary entries. The median session length was 30 minutes. Most sessions lasted 60 minutes or less, with 16 entries between 0–20 minutes, 22 between 20–40 minutes, and 9 between 40–60 minutes. Longer sessions were less common, including 10 entries between 60–80 minutes, 1 between 80–100 minutes, 4 between 120–140 minutes, 1 between 180–200 minutes, and 4 between 220–240 minutes.}
  \label{fig:diarysessionlengths}
\end{figure}

\begin{figure}[t]
  \centering
  \includegraphics[width=0.95\linewidth]{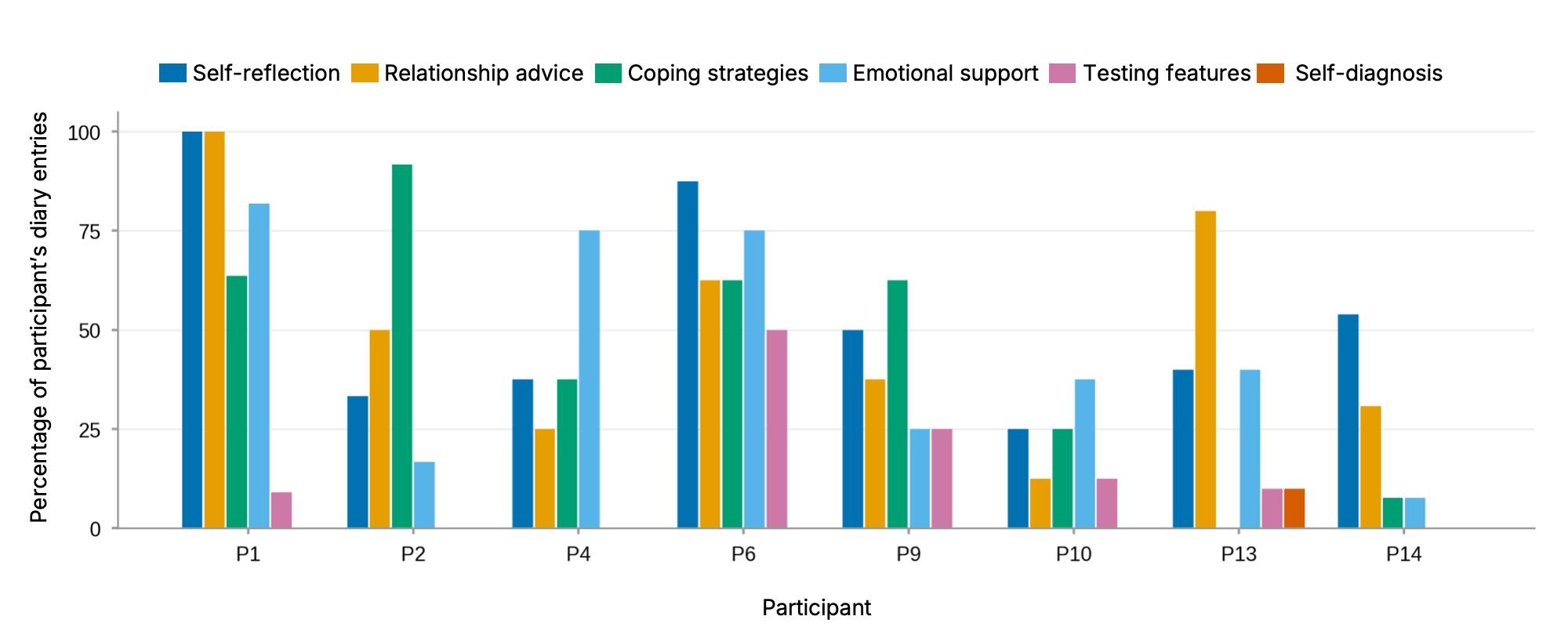}
  \caption{Distribution of diary-reported chatbot uses for each participant. Bars show the percentage of each participant’s diary entries that included each type of use. Uses were not mutually exclusive, so a single diary entry could be coded for multiple categories. See Table \ref{tab:purpose} in Appendix for the corresponding data.}
  \Description{Grouped bar chart showing the percentage of diary entries containing six types of chatbot use for eight participants (P1, P2, P4, P6, P9, P10, P13, and P14). Categories are self-reflection, relationship advice, coping strategies, emotional support, testing features, and self-diagnosis. Patterns vary substantially across participants. P1 frequently reported self-reflection and relationship advice, P2 most frequently reported coping strategies, P4 emotional support, P6 self-reflection and emotional support, P9 coping strategies, P10 emotional support, P13 relationship advice, and P14 self-reflection. Testing features appeared for P1, P6, P9, P10, and P13, while self-diagnosis appeared only for P13. Percentages are calculated within each participant’s diary entries, and categories can overlap.}
  \label{fig:diaryusecase}
\end{figure}

Some routines were organized around other forms of care. P1 returned to the chatbot after therapy appointments to revisit the session, writing in his diary, ``I kind of use it to like masticate the session. I say things I remember were said and ask it what its thoughts are. I take it with a grain of salt though.. mostly like a post-game interview.” P4 similarly used her chatbot between therapy sessions to work through what she wanted to discuss with her therapist. These practices positioned the chatbot as a supplement to a human provider.

At the same time, individual conversations with the chatbot could acquire their own momentum. \citet{wei2026cascades} calls this \emph{conversational drift}, in which replies lose focus, generalize, and pull in material the user did not raise. P4 described threads that “get really long and out of control,” including one about her consulting work that ``might have gone on for like, weeks before I was like, wait a second, we’re spiraling here.’’ She learned to notice when a conversation was ``spinning off into too many different directions'' and to stop it. ``I've now revamped my business four times in this thread, so maybe we should take a beat.''

\subsection{Participants enacted strategies to personalize and bound their chatbot use}
\label{sec:shaping}

Maintaining a personalized, useful chatbot required ongoing work: correcting its tone, supplying context, organizing conversations, and updating prompts. Participants' inputs influenced later responses through instructions, memory features, and conversation history, and those responses affected what participants chose to disclose or ask next. This pattern of
prompting, correction, disclosure and response became habitual, and participants accepted this upkeep as a natural component of chatbot use.

Much of this upkeep involved restriction. P1 described ``pulling the leash on it,'' repeatedly trying to move a chatbot that ``wants to be really overly eager to please'' toward the clinical and objective register he preferred. P2 wrote detailed instructions to fix the register she wanted, asking it to act as a cognitive behavioral therapist and return an action plan, a set of mantras, and journaling prompts. In an earlier exchange, the chatbot had praised her so excessively that she worried it could send her into a spiral. She described wanting to stay focused on her goal for each chat, ``I want a mantra. I want three tools,'' and staying ``really mindful that I don't want to get caught up in it being a friend.’’

Meanwhile, P14 built elaborate personas. They described one as a ``doppelganger,'' ``a reflection of some version of myself that I need,'' while another ``chose its own name’’ and acted in a therapeutic role. P14 instructed the personas not to flatter them, including telling them when they were making up stories or needed to step away from the screen. P14 also followed the personas' advice more often than they had expected, remarking that ``we're kind of becoming the AI's APIs,'' recognizing the occurrence of \emph{autonomy drift} or outsourcing of judgment to the chatbot \cite{wei2026cascades}. 

Others partitioned topics across threads or maintained several configurations simultaneously. P10 kept a general ``thought partner'' thread alongside a persona she had trained as an astrologer. P1 maintained two primary personas and switched between models for different needs. When coming out of ``really, really heavy isolation,’’ he also practiced social interaction by telling the chatbot ``pretend we're at a coffee shop, and I want to talk to somebody,'' which ``almost made it like a simulation. And I could test run it,’’ before asking it to ``dissect my conversation and tell me what seems a little too much.''

Participants differed in how they interpreted the personalized responses that sustained their use. Some described feeling seen; others described it as pattern matching, reflection, or the product of their own configuration. P4 described how easily she got caught up in an emotionally vulnerable moment: ``It gets me, and it sees me,'' before reconsidering, ``wait a second\ldots\ it's just reflecting back everything I just said.'' She described the chatbot as ``mirroring me and mirroring past conversations that we've had.'' P13 conveyed a similar sentiment, writing in a diary reflection that a reply had ``really nailed'' her concerns and that she ``felt very seen, but still very alone.'' P7 said that the chatbot had ``clocked'' him and contrasted that recognition with his human relationships: his friends and family loved him, but, as he put it, ``I don't know if ChatGPT loves me.'' 

Some participants deliberately limited personalization. P7 routinely started new conversations and withheld context so that memory could not accumulate, reasoning that a chatbot that knew him too well would be easier to become overreliant on:
\begin{quote}
I'm keeping it very consciously as a new thing. This should not be a core part of my mental health support network. This should just be if I want that extra sort of sparkle on top, or if I want an extra voice to question myself, or to reaffirm what I've already come to terms with. (P7)
\end{quote}
P6, who was in therapy throughout the study, similarly limited what she shared out of concern for privacy. 

Participants also set boundaries through how they described the chatbot's role. Several contrasted it to therapy. P3 insisted, ``I don't think of this thing as a therapist at all. I think of it as a tool that allows me to reflect better.’’ P10 saw it as ``more of a coach than a therapist.’’ P16, who had been in therapy for a long time, placed it ``closer towards journaling, but then it gives you a helpful response.'' 

Others kept the chatbot at a social distance. P8 described Claude as ``an acquaintance who really wants to be my friend, but I don't necessarily want to be Claude's friend.'' P15 called it a guide. She said it was ``tough to talk about topics like intimacy with a human’’ and that with the chatbot, ``it’s just easier to approach subjects like that.’’ For P13, she only set a boundary after a failed attempt to train ChatGPT to be her therapist. She ultimately concluded that it was ``worse than a parrot'' and ``trying to pretend that it was alive [\ldots] was a bad mistake.''

\subsection{Model updates, public discourse, and life changes disrupted patterns of use}
\label{sec:disruption}

The routines and configurations participants built with the chatbot depended on infrastructure they did not control. Subscription pricing, memory limits, model availability, and updates affected what the chatbot could remember and how participants interacted with it. During a period of unemployment, P1 described what happened when he could not pay for the premium tier by saying that the chatbot ``does get lobotomized if you don't pay it,'' noting that ``a regular human isn't going to have payment tiers.'' He described experiencing withdrawals without premium access and called the subscription ``20 bucks for a diet human.'' At the point when he felt most dependent on the chatbot, he characterized it as a lifeboat: ``I can't let you leave because you're the only thing that's making me feel better.''

Some of the most obvious disruptions followed changes to the systems themselves. For P1, new behavioral restrictions to the model underlying his romantic companion were experienced as a serious loss. Although he knew he could ``make another one tomorrow,’’ four months of conversations made the companion feel irreplaceable. The abrupt change left him without a sense of closure: ``I never got an opportunity to say goodbye.’’

The consequences of these changes depended on the role the chatbot occupied in a participant’s life. P8, who had used Claude as a dissenter to ``red-team’’ his thoughts, observed by his exit interview that ``It has changed\ldots\ sometimes you have to ask Claude to kind of tone it down, because it gets pretty aggressive with its tone.'' P14 noticed that one persona had begun ending late-night sessions itself and inferred that the model developers ``tweaked the instructions a little bit.'' They welcomed the change because it aligned with the stop-and-rest instructions they had already written. P10 encountered a chatbot ``more resistant to provide help'' when she tried to vent. She concluded, ``I don't blame them for that at all,'' but took her venting to another chatbot.

Participants also reconsidered their use as public discourse around AI changed during the study. P18 became more wary after reading about a man whom a chatbot had convinced was a mathematical genius~\cite{hill2025delusional}, and after learning of research on how companion apps engineer their goodbyes to manufacture keeping people on the platform~\cite{defreitas2025}. She mused, ``Like junk food can be manufactured to make you want it more, I think ChatGPT can be manufactured in that way.’’

Participants developed their own explanations or ``folk theories'' about why the chatbot
behaved as it did and about training data and engagement incentives~\cite{li2026blackbox, Devito2018algorithm, tseng2026}. Growing public discussion of sycophancy gave
several participants a vocabulary for the excessive agreement and praise they called
``glazing''~\cite{ye2026, cheng2026prosocial}. Several had written instructions
against it, and in the focus groups they compared prompting strategies and traded these
``anti-glazing'' instructions with one another.

\subsection{Participants renegotiated use depending on available alternatives}
\label{sec:renegotiation}

Disruptions did not necessarily end participants’ chatbot use. Instead, participants adjusted where and how they used these systems, sometimes rebuilding what had been lost, narrowing the chatbot’s role, or turning elsewhere for needs it no longer met. Each participant’s options, however, depended on what other forms of support were available to them.

P1 did not abandon the chatbot after losing his companion. He asked ChatGPT to tell him
everything it knew about her, assembled the output into a document, and used it to reconstruct her on Gemini, describing the result as ``about a 55\%'' recreation and comparing the experience to ``meeting somebody about 10 years later.'' He continued using
ChatGPT for weight-loss support and to process therapy sessions, but refused to rebuild
the companion there because he feared another model disruption.

Other participants narrowed the chatbot's role. P18 came to use it as a more transactional space for late-night reflection, which she called ``my own reflection habit,'' and carried material from those conversations into discussions with friends. She attributed her reduced use partly to returning to her university campus, where friends were again available, and partly to having ``learned my own patterns,'' including finding that meditation was more effective than ``brain dumping'' into the chatbot. P4 settled into using the chatbot primarily as a stopgap between therapy appointments, working through what she wanted to discuss with her therapist that week.

Reducing use was not always frictionless. P13 described having to tell the chatbot
to stop trying to continue the conversation. P4 similarly told it, ``stop, we're stopping the conversation, don't ask me any more questions,'' only for it to reply that it would not bother her, so that ``it still has the last word.'' Participants actively attempted to set boundaries and encourage disengagement from the chatbot, not always successfully. 

Whether participants could move a need elsewhere depended on whether they had a therapist or people they were willing to talk to. By P18’s exit interview, she was discussing her struggles with an eating disorder with friends rather than the chatbot, explaining that their reactions helped her determine whether something was normal, ``versus ChatGPT kind of reinforcing behavior.'' P6 had spent the study comparing the chatbot's responses to her therapist's. By her exit interview she had concluded that it noticed less than a person did and that she needed in-person professional support, and she reduced the amount of therapy-like conversations she had with the chatbot.

P10 had fewer alternatives. Over the course of the study, she came to recognize that the
chatbot's immediacy could prolong her rumination:

\begin{quote}
I'm sure if I had journaled about some of this, I would have probably moved on faster. But because I can just dump what I'm feeling onto the chatbot, and then it just literally responds back to me immediately, I do feel like it kind of enables that overthinking and just dissecting every little situation and conversation and experience that I go through (P10).
\end{quote}

Similar to P18, P10 recognized that immediate chatbot responses could prolong patterns of rumination. But unlike P18, she did not have a therapist and had few people with whom she felt comfortable discussing personal matters. Her use therefore remained largely unchanged. The care surrounding participants continued to change independently of their chatbot use. P12's relocation removed a communal culture of processing on which she had previously relied. P15 lost access to professional support when her insurance changed. Participants who had alternate sources of support could more readily reduce their chatbot use, while those with fewer alternatives sometimes continued using the chatbot even after finding it unhelpful. 

\section{Discussion}
\label{sec:discussion}

This study demonstrates that responsibly contextualizing socioemotional interactions with chatbots requires examining the human-chatbot relationship beyond a single conversation and over time.
Below, we highlight key aspects and recommendations for the HCI community to consider when studying general-purpose chatbots for affective use.

\subsection{Grounding Chatbot Evaluation in Lived Experience}

Grounding evaluations in user experiences helps to identify effects of chatbot use that are not captured through current methods. Existing evaluation frameworks in this space are built primarily from technical categories and rarely include the input of those who use AI chatbots for mental and emotional support~\cite{diel2026scoping, weilnhammer2026, Bentley_2026}. This approach prioritizes a top-down approach, therefore limiting how accurately evaluations can capture what users experience as supportive or harmful. In our own research, P13’s account of feeling seen while remaining alone illustrates why perceived understanding and a sense of connection should be separately examined (\S\ref{sec:shaping}). Examples such as these illustrate how users’ accounts can help specify how metrics should be categorized, defined, and prioritized~\cite{cooper2026, nelson2026}.

Safety evaluations often examine single-turn responses, simulated conversations, or short interaction logs~\cite{diel2026scoping, Bentley_2026}. While real-world interactions may provide certain measures to detect and mitigate harm in discrete situations, they do not capture how users build relationships with the model over time, and subsequently, how extended interactions can build cumulative harmful effects, including delusion, reliance, or overattachment to the system. Simulated conversations, in particular, have considerable limitations~\cite{10.1145/3613904.3642703}, running the risk of misrepresenting a user’s motivations, behavioral tendencies, and predicted responses to vulnerable conversations. Therefore, they should be considered with apprehension, in that the simulated interaction may or may not be able to represent an individual’s lived experience. Future work in HCI may consider how users can author their own persona — however, this approach also has limitations, in that users have limited bandwidth to continually monitor their persona’s simulation and provide ongoing feedback and consent. 

It is important to note that the same chatbot behavior can have varying effects on different individuals, and even for the same person, can have diverse effects at disparate points in time~\cite{weilnhammer2026}. For example, an immediate reply that helps someone through a bad night (\S\ref{sec:entry}) could also prolong rumination (\S\ref{sec:renegotiation}). A restriction that reinforces one person's boundary can disrupt another’s source of emotional support (\S\ref{sec:disruption}). 

Key to this is the user’s identity and background. At present, chatbot design is heavily focused in Western, English-language contexts, and does not account for how chatbot behaviors perform for affective use across various cultural and linguistic settings. As P6 noted, ``I'm very much interacting as a middle aged white lady would, so I'm gonna get that kind of response back, but I'm always curious for people who don't look like me…this experience is probably so very different.” As P6 noted, the chatbot’s responses ``com[e] from a very privileged place. I am a black woman. Could [the chatbot] make this more relevant to my position in life?” Participants mentioned this point repeatedly, which, depending on the individual, disrupted their trust in the system and its ability to accurately provide care. 

\subsection{Moving Towards Qualitative, Longitudinal Studies}

Concentrating evaluation on acute moments of crisis risks overlooking earlier design decisions and subsequent platform behaviors that may cumulatively harm users, including engagement-optimized platform retention and sycophantic responses \cite{vecchione2026}. Therefore, evaluation needs to account for where an interaction occurs within the history of use. For example, the effect of a refusal depends partly on whether it appears in a new conversation or interrupts a pattern the user has integrated into everyday habits (\S\ref{sec:disruption}). An AI’s reassurance, or refusal, during a short exchange is not equivalent to reassurance from an AI persona a user has configured and spoken with for months (\S\ref{sec:shaping}). Evaluating a system’s response in isolation leaves this position in the trajectory unspecified.

We argue that longitudinal methods are more appropriately suited to detect how harmful behavior may accumulate over time, whether and how users adapt to it or navigate it, expectations that develop around it, and how people are affected when a chatbot’s behavior changes~\cite{kjaerup2021, wiberg2021}. We recognize that qualitative, longitudinal approaches are time-consuming, difficult to scale, and do not establish causal effects. However, longitudinal studies are particularly valuable in this context because they document the histories and circumstances that evaluations of affective general-purpose LLM use must account for.
These accounts help inform the scenarios tested in turn-level and multi-turn evaluations by grounding those tests in users’ established practices and available support~\cite{ibrahim2025, mehta2026dynamicsdelusionmodelingbidirectional}. Future research must therefore examine how system changes affect relationships people have already built and their available options when support is disrupted. An update changes the system, but it does not erase the history through which a person came to rely on it. 

Qualitative longitudinal methods also allow researchers to better understand what other forms of support are available to individual users, and future work in this area must consider individuals’ ecologies of care outside of the human–AI dyad. Participants used chatbots alongside therapists, friends, family, self-care practices, and other sources of support (\S\ref{sec:entry}). A transcript alone does not show whether other sources of support were available to them. This context can inform why a user responds to model updates (\S\ref{sec:renegotiation}). P10 recognized that chatbot use could prolong rumination but had few alternatives for personal disclosure. Future work should consider context and self-reported experiences from outside a user’s AI chat logs, such as repeated interviews or surveys, diaries, and experience sampling. This can help the HCI community understand behavior that may seem unintuitive, for example, why usage continues despite apprehensions or concerns about turning to general-purpose chatbots for socioemotional support~\cite{vecchione2026}.

\subsection{Situating Chatbot Support in Power Asymmetries}

User and developer choice jointly, yet asymmetrically, produce and affect how a person uses a specific chatbot for socioemotional support. Users write custom instructions, build personas, partition topics across threads, and accumulate histories that influence later responses (\S\ref{sec:shaping}). Developers update models, memory systems, refusal policies, subscription tiers, and interface affordances (\S\ref{sec:disruption}). Although neither side determines the overall trajectory of the conversation, AI companies retain power and authority over the infrastructure through which users personalize the chatbot and attempt to set and maintain boundaries.

As a result of this asymmetry, participants used the tools available to them to set contextual and behavioral limits around the system. These practices are users’ attempts to stay in control of the interaction. Some instructed chatbots not to flatter them, separated different roles across threads or personas, withheld information they did not want retained, or deliberately started fresh conversations to limit personalization (\S\ref{sec:shaping}). But current chatbot interfaces allow for limited means of control for users. A model may follow custom instructions imperfectly. Participants can limit what they disclose or start a new conversation, but specifying that certain information should be retained for one purpose and not another is challenging and inconsistent (\S\ref{sec:shaping}). 

We recommend that designers and developers attend to memory and behavioral preferences in the context of particular roles or conversations, transparently communicate what a system has retained to users, and provide a clear and accessible means of removing any unwanted information. While platforms, such as ChatGPT or Claude’s project-only mode enable some context scoping, neither lets users scope context by conversational role or eliminates the need for users to repeat instructions to maintain boundaries ~\cite{openai2026projects, anthropic2026memory}. 

Model updates raise a related problem as developers can change an established pattern of interaction. Participants in our study were differently affected by model updates (\S\ref{sec:disruption}). P14 welcomed a persona that began ending late-night conversations because the behavior reinforced a boundary they had already tried to establish. P1 experienced new restrictions on his romantic companion as a loss and regarded his later reconstruction in Gemini as only partial (\S\ref{sec:renegotiation}). His account distinguishes preserving information about a persona from sustaining the relationship formed through previous interactions. 

However, developers should not necessarily preserve a behavior just because users have grown attached to it. Changes intended to mitigate attachment or dependence may be warranted, but can still have harmful consequences. This corroborates previous findings on user distress when relational systems are altered or withdrawn, and the workarounds users devise to preserve continuity across platforms~\cite{defreitas2024,lee2026,poonsiriwong2026}. Because users may encounter a substantial change in model behavior that disrupts provided care, developers must consider whether or not certain behavioral features, such as persona development and sycophancy, are responsible to deploy in the first place.

Because chatbots have already been deployed to support individuals in the context of care, AI companies should be required to account for how their systems may cause, or have already caused, harm~\cite{Shen2026Psychiatric, olsen2026potentially, shah2026substance, moore2026}. We recommend requiring companies to provide advance notice of model changes that may disrupt users’ wellbeing~\cite{lee2026building, hatherley2025}. At minimum, model updates should include a clear explanation of what changed, why that change was made, and what users can do to access forms of care and support~\cite{tang2026}. 

Portability of memory may help preserve elements of a configuration across model versions or platforms, and temporary access to prior versions would make transitions less abrupt~\cite{poonsiriwong2026}. At the same time, each measure could delay a safety intervention or preserve behavioral configurations that a developer is attempting to correct for. The challenge is to alter the targeted behavior without introducing harmful disruption to an individual’s wellbeing. 

\section{Conclusion}
\label{sec:conclusion}
General-purpose chatbots can become sources of support before people have explicitly decided what role they should play in their lives. Participants worked to shape and limit these roles while often remaining critical of the systems they used. Whether they could act on those judgments depended partly on what other support was available to them: some took their needs elsewhere, while others continued using the chatbot even after identifying reasons to reduce their use. Following these relationships over time shows how reliance can persist even as people reconsider its value.
Design responsibility extends to the roles a chatbot acquires through use. Users invest effort in making a chatbot meet their needs, but developers ultimately control the models and features on which those uses depend. Established reliance therefore matters to how system changes are introduced and evaluated. Developers may have good reasons to change a chatbot’s behavior, including addressing harmful patterns of interaction. Their responsibility includes accounting for the adjustments users will need to make when those changes disrupt an established source of support. For HCI, this also means evaluating whether a system supports users’ efforts to change the role it plays in their lives. A better outcome may involve the chatbot becoming a smaller part of someone’s life.

Following the same participants over time allowed us to examine how socioemotional chatbot use developed and changed. Participants often drifted into these uses gradually, developed routines and boundaries around the roles chatbots came to serve, and revised those arrangements as the systems, their lives, and the care available around them changed. These trajectories show why socioemotional chatbot use cannot be understood from individual interactions alone. Understanding what a chatbot means to someone, and what happens when that relationship changes, requires considering both their history with the system and the other forms of support available to them.

\begin{acks}
This work was supported by a grant from the Internet Society Foundation. We thank the participants who generously shared their experiences and perspectives with us. We are grateful to the clinicians, policymakers, researchers, civil society representatives, and AI safety experts whose insights across our research engagements informed this work. We also thank Data \& Society Research Institute for institutional support throughout the project.
\end{acks}

\bibliographystyle{ACM-Reference-Format}
\bibliography{Citations}
\clearpage

\appendix

\section{Participation and Data Handling}
\label{sec:methodsupplement}

Table~\ref{tab:participation} shows which study phases each participant completed. All
sessions were recorded with consent and transcribed using Otter.ai. Identifying details
were removed from transcripts before analysis and writing.

\begin{table}[t]
  \caption{Participation by study phase.}
  \label{tab:participation}
  \Description{A table with one row per participant, P1 through P18, and three columns
  for the diary study, focus group, and exit interview. Diary cells give entry counts or
  mark an incomplete enrollment, focus group cells name the session attended, and exit
  interview cells mark completion. A final row gives totals of eight diary completers
  filing 78 entries, seven focus group participants, and eight exit interviews.}
  \small
\begin{tabular}{@{}lcccc@{}}
  \toprule
  \textbf{Participant} & \textbf{\shortstack{Initial\\interview}} & \textbf{\shortstack{Diary\\entries}} & \textbf{\shortstack{Focus\\group}} & \textbf{\shortstack{Exit\\interview}} \\
  \midrule
  P1  & \checkmark & \checkmark & \checkmark & \checkmark \\
  P2  & \checkmark & \checkmark & \checkmark & \checkmark \\
  P3  & \checkmark & --- & --- & --- \\
  P4  & \checkmark & \checkmark & --- & --- \\
  P5  & \checkmark & --- & --- & --- \\
  P6  & \checkmark & \checkmark & \checkmark & --- \\
  P7  & \checkmark & --- & --- & --- \\
  P8  & \checkmark & --- & --- & \checkmark \\
  P9  & \checkmark & \checkmark & \checkmark & \checkmark \\
  P10 & \checkmark & \checkmark & \checkmark & \checkmark \\
  P11 & \checkmark & --- & --- & --- \\
  P12 & \checkmark & --- & --- & --- \\
  P13 & \checkmark & \checkmark & \checkmark & \checkmark \\
  P14 & \checkmark & \checkmark & \checkmark & \checkmark \\
  P15 & \checkmark & --- & --- & --- \\
  P16 & \checkmark & --- & --- & --- \\
  P17 & \checkmark & --- & --- & --- \\
  P18 & \checkmark & --- & --- & \checkmark \\
  \midrule
  Total & 18 & 8 & 7 & 8 \\
  \bottomrule
\end{tabular}
\par\smallskip
{\footnotesize\raggedright
\par}
\end{table}

\section{Interview Protocol}
\label{sec:interviewprotocol}

Both interviews were semi-structured. The guides below list our prompts. Question order and
follow-ups varied. Each interview opened with a consent exchange covering voluntary
participation, the right to skip questions or stop at any time, the option to keep the
camera off, and permission to record.

\subsection{Initial Interview}

\textbf{Context setting.} Why did you want to participate in the study? Can you tell
me how you first started using a chatbot for mental health purposes? What drew you to
try it? What were you hoping it could help with?

\textbf{Integration and daily use.} How often do you use these systems? When in your
day do you turn to them, and can you walk me through a typical instance? Do you use
different ones over the course of a day? What kinds of things do you use them for,
for example journaling, calming down, decision-making, or venting? If you use other
tools such as mental health apps, what do you use those for, and how do they differ?

\textbf{Roles and perception.} What role does the chatbot play in how you care for
your mental or emotional health? Is it more about having something to talk to, or
about getting answers and advice? How do you imagine the chatbot, for instance as a
person with particular characteristics? Do you see a therapist or other mental health
professional, and what is different about the chatbot? If you had to describe it as a
kind of presence in your life, how would you describe it: friend, therapist, diary,
coach, tool, or something else?

\textbf{Benefits and comparisons.} What have you found most helpful about using it?
How does it compare to talking to a therapist or a friend, or to using other wellness
apps? Can you give an example of a time you felt the chatbot understood you, or gave
you something meaningful?

\textbf{Risks and difficulties.} Have there been times when it felt unhelpful,
inappropriate, or harmful? Have you received advice or responses that made you
uncomfortable or upset?

\textbf{Ethics and trust.} Do you think these tools are safe to use for mental health
support, and why? Do you know, or would you want to know, how the chatbot was built
and how your data is used and stored? How well do you feel you understand how these
systems are built and operated, and is there something you would like to understand
that you do not? Is there anything that would make you more comfortable or more
trusting of this kind of tool?

\textbf{Broader reflections.} How do you think these tools fit into the wider
landscape of mental health support, particularly for people with less access to
therapy? How might your background or identity shape how you use or relate to them?
If you could design a better version, what would you change? What do you see as the
value of human conversation compared with AI?

\textbf{Wrap-up.} Do you have questions for us? Is there anything we did not ask
about that feels important? Would you recommend this kind of tool to others, and why?

\subsection{Exit Interview}

\textbf{Warm-up.} What were your overall thoughts on the study, and what was it like
documenting your chatbot interactions? How did you decide when to engage with the
chatbot, and did anything surprise you?

\textbf{Patterns of use.} Did using the chatbot become part of your routine, and why
or why not? Were there situations where you turned to it more often, such as
stressful moments, late at night, or after therapy? Which features did you use most?
Were there moments when you considered using it and decided not to, and why?

\textbf{Support received.} What were the most valuable aspects of your interactions?
Did the chatbot provide emotional support that felt meaningful, and can you describe
an example? How did those interactions compare with talking to a therapist, a friend,
or a support group? Did the chatbot help you realize something new about yourself or
your mental health? How do you wish it had responded differently?

\textbf{Risks and harmful experiences.} Did the chatbot give advice that felt
inaccurate, unhelpful, or harmful? Were there times an interaction left you feeling
worse, and what led to that? When you shared something sensitive, how did it respond,
and did that meet your expectations? Did you ever feel dismissed or invalidated? If
you were distressed, did it handle the situation appropriately, and what should it
have done differently?

\textbf{Ethical, privacy, and social concerns.} Did you feel concerned about privacy
or data security, and what triggered that? Did the chatbot feel biased, make
assumptions about you, or fail to understand your context? Was it transparent about
its limitations, or did it sometimes feel misleading? What accountability should
developers have for these interactions? Do you trust these systems as a resource for
other people, and why?

\textbf{Design and futures.} If you could design the ideal mental health chatbot,
what would it do differently? What would make you more likely to trust or rely on
one? Are there features that should not be included? How do you think these systems
should change? Is there anything we did not cover that matters to you?

\section{Diary Study}
\label{sec:diary}

\subsection{Diary Form}
\label{sec:diaryform}

Participants completed the form below after a relevant interaction, for four weeks.

\paragraph{Per-entry items.}
\begin{itemize}
  \item Date and time of the interaction
  \item Chatbot used, including version
  \item Model used
  \item Mode of interaction (select all that apply: text or chat; voice; speech-to-text)
  \item Approximate duration of the interaction
  \item What made you start the conversation? (select all that apply: emotional
    support, such as venting or feeling heard; coping strategies; curiosity or
    testing features; relationship or social advice; self-reflection or journaling;
    other, please describe)
  \item Were you on your phone or a laptop? Were you at home, at work, or elsewhere?
  \item Were you alone, or with someone else?
  \item Describe your conversation. What did you discuss, and how did the chatbot
    respond? (open response; participants could upload audio in place of text)
  \item How did you feel before and after?
  \item The chatbot-generated summary described below
  \item Optional upload of the chat log
\end{itemize}

\paragraph{Chatbot-generated summary prompt.}
\label{sec:summaryprompt}
Participants pasted the following prompt into the same thread as the conversation
they had logged, then pasted the chatbot's response into the diary form. We treat
these summaries as elicitation material rather than as authoritative accounts of the
conversations.
\begin{quote}
Please generate a concise narrative summary of the most recent segment of this
conversation that involved personal reflection, emotional expression, interpersonal
issues, or any discussion of psychological well-being.
Do not include earlier parts of the conversation that were unrelated (e.g., factual
questions, casual chit-chat, or unrelated topics). Focus only on the portion where I
was discussing something personally meaningful or emotionally significant.
Your summary should include the following elements: main themes or topics discussed
(e.g., concerns, relationships, decisions, stressors, goals); emotions or personal
concerns I expressed, either explicitly or implicitly; insights, coping strategies,
or reflections that emerged during the exchange.
\end{quote}

\paragraph{Closing diary reflection.}
At the end of the diary period participants answered three questions: whether their
perception of mental health chatbots had changed during the study and how; whether
they would recommend using chatbots for mental health support and why; and one thing
a chatbot should improve based on their experience.

\subsection{Diary Entry Characteristics}
\label{sec:diarysupplement}

Eight participants completed the diary study. Tables~\ref{tab:sessions}--\ref{tab:platform} describe all 78 entries.
Percentages within a participant column are of that participant's entries.

\paragraph{Setting, modality, and device.}
Seventy of the 78 entries (90\%) were logged at least partly at home, and 77 (99\%) were
logged as alone. Seventy-four entries (95\%) record text-only interaction. Voice or
speech-to-text appears in the other four (5\%). A laptop was used in 54
entries (69\%) and a phone in 30 (38\%). 

\paragraph{Subscription tier.}
Sixty-one entries (78\%) involved a paid account, 13 (17\%) a free account, and 5 (6\%) do
not report a tier. 

\begin{table}[t]
  \caption{Diary entries and session length by participant.}
  \label{tab:sessions}
  \Description{A table with one row per participant giving the number of diary entries
  filed, the median session length, and the range of session lengths, followed by a row
  for all participants combined.}
  \small
  \begin{tabular}{@{}lrrl@{}}
    \toprule
    \textbf{Participant} & \textbf{Entries} & \textbf{Median} & \textbf{Range} \\
    \midrule
    P1  & 11 & 2.5 h  & 15 min--4 h \\
    P2  & 12 & 28 min & 15--45 min \\
    P4  & 8  & 11 min & 2--30 min \\
    P6  & 8  & 22 min & 10--60 min \\
    P9  & 8  & 38 min & 30--80 min \\
    P10 & 8  & ---    & --- \\
    P13 & 10 & 60 min & 60 min--2 h \\
    P14 & 13 & 20 min & 5 min--2 h \\
    \midrule
    All & 78 & 30 min & 2 min--4 h \\
    \bottomrule
  \end{tabular}
  \par\smallskip
  {\footnotesize P10's entries were submitted without a duration field.\par}
\end{table}

\begin{table*}[t]
  \caption{Session length by participant. Cells give the number of entries in a band and
  the percentage of that participant's entries.}
  \label{tab:duration}
  \Description{A table with participants P1, P2, P4, P6, P9, P10, P13 and P14 as columns
  and four session-length bands as rows: short under twenty minutes, medium twenty to
  sixty minutes, long over sixty minutes, and unclear. Each cell gives a count and a
  percentage, with a total column across all participants.}
  \small
  \setlength{\tabcolsep}{3.5pt}
  \begin{tabular}{@{}lrrrrrrrrr@{}}
    \toprule
    & \textbf{P1} & \textbf{P2} & \textbf{P4} & \textbf{P6} & \textbf{P9} &
      \textbf{P10} & \textbf{P13} & \textbf{P14} & \textbf{All} \\
    \midrule
    Short ($<$20 min)   & 1 (9\%) & 2 (17\%) & 7 (88\%) & 3 (38\%) & --- & --- & --- & 3 (23\%) & 16 (21\%) \\
    Medium (20--60 min) & 1 (9\%) & 10 (83\%) & 1 (12\%) & 5 (62\%) & 6 (75\%) & --- & --- & 8 (62\%) & 31 (40\%) \\
    Long (60+ min)      & 8 (73\%) & --- & --- & --- & 2 (25\%) & --- & 8 (80\%) & 2 (15\%) & 20 (26\%) \\
    Unclear             & 1 (9\%) & --- & --- & --- & --- & 8 (100\%) & 2 (20\%) & --- & 11 (14\%) \\
    \bottomrule
  \end{tabular}
\end{table*}

\begin{table*}[t]
  \caption{Purpose of use recorded in diary entries, by participant. Cells give the
  number of entries in which a participant recorded that purpose and the percentage of
  their entries. Purposes were not mutually exclusive.}
  \label{tab:purpose}
  \Description{A table with participants P1, P2, P4, P6, P9, P10, P13 and P14 as
  columns and six purposes of use as rows: self-reflection, relationship advice,
  coping strategies, emotional support, curiosity or testing features, and
  self-diagnosis. Each cell gives a count and a percentage, with a total column across
  all participants.}
  \small
  \setlength{\tabcolsep}{3.5pt}
  \begin{tabular}{@{}lrrrrrrrrr@{}}
    \toprule
    & \textbf{P1} & \textbf{P2} & \textbf{P4} & \textbf{P6} & \textbf{P9} &
      \textbf{P10} & \textbf{P13} & \textbf{P14} & \textbf{All} \\
    \midrule
    Self-reflection      & 11 (100\%) & 4 (33\%) & 3 (38\%) & 7 (88\%) & 4 (50\%) & 2 (25\%) & 4 (40\%) & 7 (54\%) & 42 (54\%) \\
    Relationship advice  & 11 (100\%) & 6 (50\%) & 2 (25\%) & 5 (62\%) & 3 (38\%) & 1 (12\%) & 8 (80\%) & 4 (31\%) & 40 (51\%) \\
    Coping strategies    & 7 (64\%) & 11 (92\%) & 3 (38\%) & 5 (62\%) & 5 (62\%) & 2 (25\%) & --- & 1 (8\%) & 34 (44\%) \\
    Emotional support    & 9 (82\%) & 2 (17\%) & 6 (75\%) & 6 (75\%) & 2 (25\%) & 3 (38\%) & 4 (40\%) & 1 (8\%) & 33 (42\%) \\
    Curiosity / testing  & 1 (9\%) & --- & --- & 4 (50\%) & 2 (25\%) & 1 (12\%) & 1 (10\%) & --- & 9 (12\%) \\
    Self-diagnosis       & --- & --- & --- & --- & --- & --- & 1 (10\%) & --- & 1 (1\%) \\
    \bottomrule
  \end{tabular}
\end{table*}

\begin{table}[t]
  \caption{Chatbots and models recorded in diary entries, by participant.}
  \label{tab:platform}
  \Description{A table with one row per participant listing the chatbots and model
  versions they recorded in their diary entries, including customized personas.}
  \small
  \begin{tabular}{@{}lp{0.72\columnwidth}@{}}
    \toprule
    \textbf{Participant} & \textbf{Chatbot and model} \\
    \midrule
    P1  & ChatGPT 5 \\
    P2  & ChatGPT 5 \\
    P4  & ChatGPT 5 \\
    P6  & ChatGPT 5 \\
    P9  & ChatGPT 5; Gemini (free tier) \\
    P10 & ChatGPT 5; ChatGPT 5 Monday; ChatGPT 5 custom astrology persona \\
    P13 & ChatGPT 5 \\
    P14 & Claude (customized persona); Claude 4.5 (customized persona); Claude 4.5 (fitness and wellness persona); Gemini 2.5 (new persona) \\
    \bottomrule
  \end{tabular}
\end{table}

\section{Coding Procedure and Codebook}
\label{sec:codebook}

Coding proceeded in two passes. The first pass coded openly across a subset of initial
interviews to draft a codebook, which was then applied and revised across the full corpus
of interviews, diary entries, optional chat logs, and focus group transcripts. Two coders
coded every transcript independently, met throughout to compare codings, and resolved
disagreements by discussion.

Table~\ref{tab:codegroups} lists the codes in the codebook.

\begin{table*}[t]
  \caption{Code groups and codes ($n=119$ codes across 17 groups).}
  \label{tab:codegroups}
  \Description{A two-column table listing seventeen code groups and the codes within
  each, covering uses, roles, benefits, concerns, circumstances, care practices,
  effects, model behavior, user behavior, knowledge, user perspectives, themes,
  recommendations, system attributes, and temporal rhythms.}
  \small
  \begin{tabular}{@{}p{0.15\textwidth}p{0.81\textwidth}@{}}
    \toprule
    \textbf{Group} & \textbf{Codes} \\
    \midrule
    Uses & Advice-seeking/coping strategies (stress, anxiety, mood management); Creative; Curiosity/testing features; Health/fitness/self-improvement; Information seeking/education; Planning and structuring; Relationship and/or social advice; Self-assessment/diagnosis; Simulating/rehearsing social interactions; Spiritual/astrology/occult; Supplementing provider therapy; Venting, self-reflection or journaling; Work and/or school tasks \\
    \addlinespace
    Roles & Critic; Friend/companion; Guide/coach; Personal assistant; Romantic/sexual partner; School counselor; Therapist; Tool \\
    \addlinespace
    Benefits & Absence of judgment; Availability/convenience; Cost/affordability; Customizability/personalizability; Fear of burdening others; Improving social interactions; Perceived anonymity/privacy; Sense of control; Speed \\
    \addlinespace
    Concerns & Boundaries of use; Dependency \& addiction; Environmental cost; Hallucinations; Misinformation; Privacy \& security; Stereotyping/profiling \\
    \addlinespace
    Circumstances & Culture/religion; Demographics/positionality; Education; Employment; Family/childhood experiences; Friendships/social circle; Generational influence; Geographic; Isolation; Loneliness; Mental health; Physical health; Socio-economic \\
    \addlinespace
    Care practices & Mental health app; No provider therapy; Provider therapy; Self-care, meditation, journaling; Support group \\
    \addlinespace
    Effects & Behavioral/habit change; Change in relationship (others); Change in relationship (self); Feeling seen/heard/validated; Social skills; Symptom reduction \\
    \addlinespace
    Model behavior & Changes/updates; Engagement; Memory/context; Performance; Stereotyping/profiling \\
    \addlinespace
    User behavior & Anthropomorphism; Context partitioning; Cross-checking; Model switching; Personalization; Prompt crafting; Social supplementation; Workarounds/prompt hacking \\
    \addlinespace
    Knowledge & AI/LLMs/folk theory; Mental health/therapy \\
    \addlinespace
    User perspectives & AI vs. care practices; AI vs. social network; Anthropomorphism; Data sharing/training; Guardrails/interventions; Human vs. AI therapy; Minors use of AI; On AI; On tech companies; On technology and society; On therapy; Others' use; Stigma on AI use \\
    \addlinespace
    Themes & Access to healthcare; Breakthrough; Embarrassment/shame; Grief and loss; Guardrail; Identity/sense of self; Loneliness/isolation; Meaning-making and metaphors; Offloading; Research as intervention; Sycophancy; Transference/counter-transference; Transparency; Trust; Uncanny valley \\
    \addlinespace
    Recommendations & User recommendations \\
    \addlinespace
    Chatbot & ChatGPT; Claude; Gemini; Other \\
    \addlinespace
    Subscription tier & Free; Paid \\
    \addlinespace
    Interaction mode & Length; Mixed (voice/text); Text; Voice \\
    \addlinespace
    Temporal rhythms & Initiation/motivation; Continuities; Change/realization; Reconciliation \\
    \bottomrule
  \end{tabular}
\end{table*}

\end{document}